# Tracking the diffusion-controlled lithiation reaction of LiMn$_2$O$_4$ by in-situ TEM

*Torben Erichsen[†], Björn Pfeiffer[†], Vladimir Roddatis[†], and Cynthia A. Volkert*[†,§]*

[†]Institute of Materials Physics, University of Göttingen, 37077 Göttingen, Germany

[§]The International Center for Advanced Studies of Energy Conversion (ICASEC), University of Göttingen, 37077 Göttingen, Germany

lithium manganese oxide spinel, tetragonal Li$_2$Mn$_2$O$_4$, EELS, twinning, defects, interface, lithiation

Spinel lithium manganese oxide (Li$_x$Mn$_2$O$_4$) is used as an active material in battery cathodes. It is a relatively inexpensive and environmentally friendly material, but suffers from capacity fade during use. The capacity losses are generally attributed to the formation of the tetragonal phase (x > 1) due to overpotentials at the surfaces of the micron-sized particles that are used in commercial electrodes. In this study, we investigate the mechanisms of tetragonal phase formation by performing electrochemical lithiation (discharging) in-situ in the transmission electron microscope (TEM) utilizing diffraction and high resolution as well as spectroscopy. We observe a sharp interface between the cubic spinel (x = 1) and the tetragonal phase (x = 2), that moves under lithium diffusion-control. The tetragonal phase forms as a complex nanotwinned microstructure, presumably to relieve the stresses due to expansion during lithiation. We propose that the twinned microstructure stabilizes the tetragonal phase, adding to capacity loss upon deep discharge.

**Introduction**

The need for lightweight power sources with high energy density is ever growing with increasing demand in all fields of application like electrified transport and temporary energy storage for alternative energy production. Many of today's secondary battery technologies involve cathodes containing significant amounts of expensive and poisonous cobalt. An alternative is transition metal based spinels that have been extensively studied over the last few decades mostly by non-local methods such as cyclic voltammetry and x-ray diffraction (XRD). However, they show problems in capacity retention, as summarized by Thackeray[1], which limits their applicability.

$Li_xMn_2O_4$ spinel (LMO), which is a low-cost and environment-friendly material, is able to accommodate lithium up to x = 2 per chemical unit while retaining the same basic arrangement of manganese-oxygen-octahedrons. In the range of 0 < x < 1, where the crystal structure remains cubic, only the lattice parameter changes with x (Fd3m, a = 8.02-8.24Å).[2] Upon increasing the lithium content above x = 1, the previously tetrahedrally coordinated lithium immediately shifts to octahedral sites while leaving the oxygen-manganese arrangement unchanged. This lowers the Mn valence below the x = 1 value of 3.5 and produces Jahn-Teller (JT) mediated tetragonal distortion.[3] The high energies of having neighboring tetrahedral and octahedral sites populated by lithium lead to low solubility between the x = 1 and x = 2 phases and to a large miscibility gap between the two phases. The atomic structures of the cubic x = 1 and tetragonal x = 2 phases are shown in **Figure 1**. The primitive tetragonal unit cell ($I4_1/amd$, a = 5.7Å , c = 9.3Å) is rotated by 45° relative to the cubic unit cell.[4] However, since the Mn and O atoms shift only slightly during the transformation, it is often simpler and more instructive in the discussion of crystallographic relationships to describe the tetragonal phase using a non-primitive tetragonal unit cell ($F4_1/ddm$, a = 8.0Å, c = 9.3Å) where the a and c axes are parallel to the ones of the cubic spinel system.[5] This has been referred to as the pseudo-cubic unit cell and vectors in this system are labeled with a subscript "p" to distinguish them from the cubic "c" and primitive tetragonal "t" systems.[6]

The known degradation mechanisms of LMO include dissolution of manganese into the electrolyte[7–10], mechanical degradation due to volume changes of the electrode particles during de-/intercalation[11,12], and capacity loss due to the fact that the cubic to tetragonal transformation at x > 1 is not entirely reversible[2]. This limits the Li uptake to half its achievable value in commercial application, since only the cubic region is used. But residual tetragonal regions have also been detected by TEM in LMO electrode particles cycled within 0 < x < 1, presumably due to the formation of the tetragonal phase in overpotentials.[11] To improve capacity retention and reversibly extend the lithium range to x > 1, a variety of costly methods have been introduced, such as using nanoscale particles for better stress accommodation[13], surface coatings[14], and partial substitution of manganese by other transition metal which reduces manganese dissolution and shifts the onset of the tetragonal transformation to larger values of x.[9] These approaches have produced moderate improvements in capacity retention, but usually the cost and environmental advantages of LMO as well as some capacity are partially or fully lost. It is somewhat surprising then that the exact mechanisms of capacity fade in micrometer-size particles due to the tetragonal transformation have not yet been fully investigated.

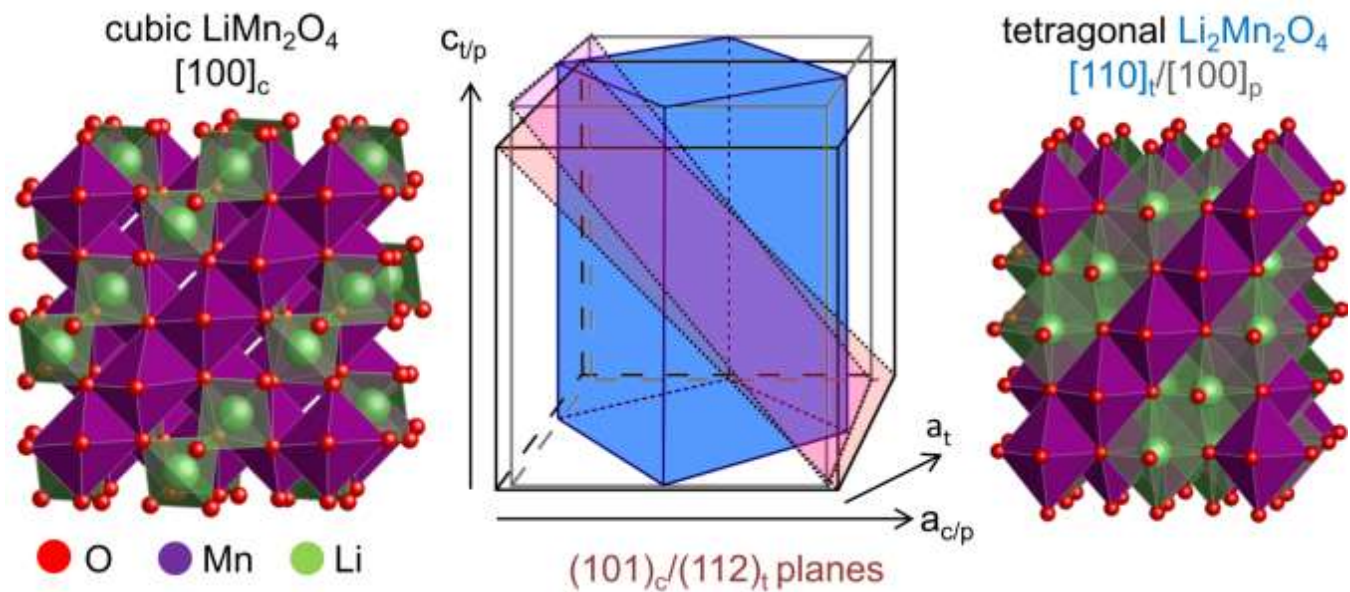

**Figure 1.** The crystal structure of $Li_xMn_2O_4$. Equivalent cuboids viewed along $[100]_c$ in cubic $LiMn_2O_4$ (left) and along $[110]_t/[100]_p$ in tetragonal $Li_2Mn_2O_4$ (right). The $MnO_6$ octahedra (violet) are distorted, but their arrangement remains unchanged upon transformation between the phases. The central sketch shows the relationship between the unit cells of the two materials, including the labeling conventions. It also shows the tetragonal twinning plane $(112)_t/(101)_p$ which lies close to the cubic $(101)_c$ plane.

LMO has so far only been studied with in-situ TEM for the case of $LiMn_2O_4$ nanowires, in which a cubic-to-tetragonal phase transformation was confirmed by electron diffraction upon lithiation.[15] But in contrast to previous reports, the transformation was fully reversible and proceeded via an intermediate orthorhombic phase. It was suggested that this could be due to particle size effects previously observed by XRD on large ensembles of LMO particles.[16] A closer study to better understand and confirm these observations requires higher spatially resolved methods such as have been applied to $LiMn_{1.5}Ni_{0.5}O_4$[17] and $LiFePO_4$[18] using x-ray methods. Comprehensive TEM-based studies including STEM-EELS (which have been applied to e.g. LiFePO$_4$[19]) could not be performed on the nanowires probably due to their thickness limiting the applicability of EELS and STEM.

As a first step in understanding how the tetragonal phase may lead to capacity fade in LMO, we track the microstructural changes during electrochemical lithiation of LMO electrode particles using in-situ TEM studies. With the help of in-situ TEM, diffraction, STEM and EELS and laser-assisted Atom Probe Tomography we are able to track the phase boundary and the associated changes in microstructure during lithiation above x>1, i.e. deep discharge. The unexpected formation of a twinned microstructure indicates an important role of lithiation-induced stresses and may stabilize the tetragonal structure against delithiation thereby providing a possible specific mechanism for capacity fade.

**Results and Discussion**

In-situ lithiation of a LMO specimen prepared by FIB from commercially available micron-sized particles was performed in the TEM by mechanically contacting the foil with a lithium tip (see Methods section for more detail). Bending contours in bright field (BF) imaging clearly reveal a reaction front moving from the contact with the Li tip into the LMO sample (e.g. **Figure 2 (a)**). The reaction front movement is accelerated by applying a -5 V bias and can be stopped during the experiment by mechanically separating the foil from the lithium source. Although the reaction proceeds without the application of an external bias, we apply a bias here to compensate for any voltages generated in the sample by electron beam illumination[20] and to overcome possible chemical potential barriers between the tip and sample.[21]

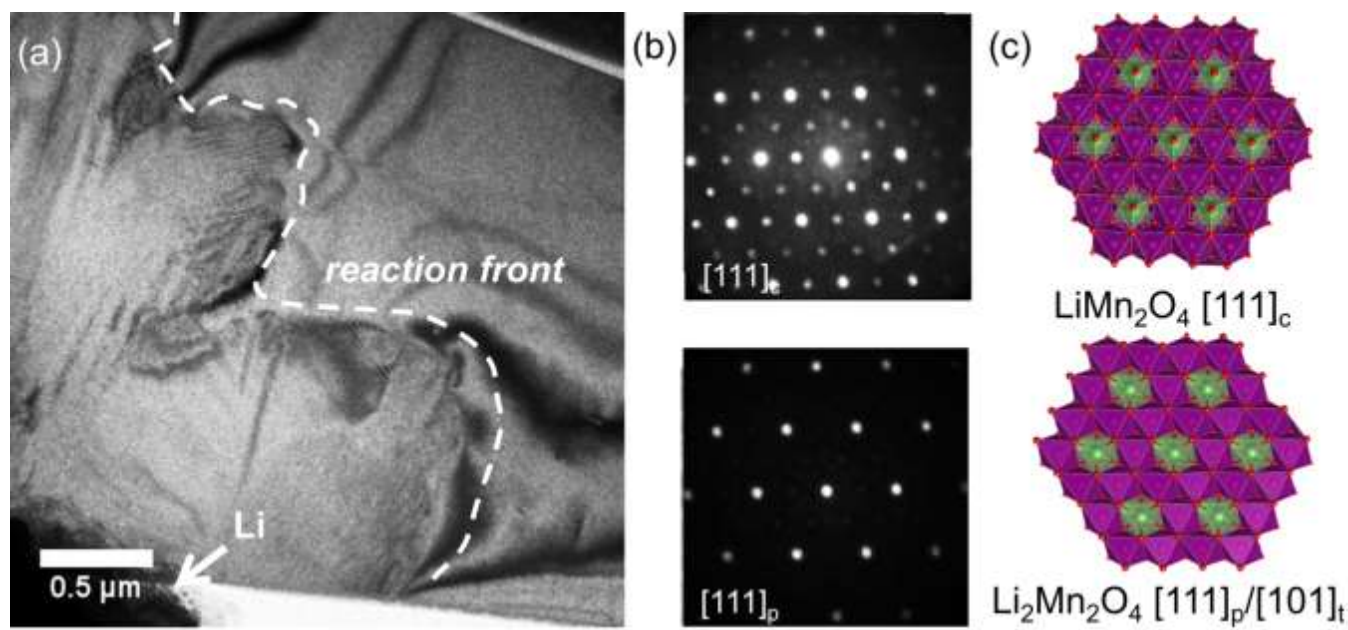

**Figure 2.** In-situ lithiation of LMO. (a) BF image of an initially single crystalline LMO specimen during lithiation. The lithium tip is visible in the lower left corner. Strain and microstructural contrast reveal a sharp interface (dashed line) between the untransformed single crystal on the top right and the lamellar structure in the transformed region on the bottom left. (b) Diffraction patterns from a scanning nanodiffraction analysis confirm the initial spinel structure in a near $[111]_c$ orientation (top) and the structural transformation to the tetragonal phase in a near $[111]_p$ orientation (bottom). (c) Sketches of the crystal structures viewed along the [111] direction show the similarity between the phases.

The portion of the sample shown in **Figure 2 (a)** is completely lithiated within 180s, which corresponds to a very high charging rate of approximately 20C. A video of this reaction is available in the Supporting Information. Bending contours are visible directly after contact with the lithium tip and move with the reaction front into the sample. The bending contours are localized directly at the interface, presumably because the phase transformation involves distortion of the original cubic spinel lattice to a tetragonal crystal structure (**Figure 1**), helping to identify the interface shape and movement.

Diffraction patterns (**Figure 2 (b)**) confirm that the sharp boundary observed in the TEM (**Figure 2 (a)**) is the phase interface between the initial $LiMn_2O_4$ and lithiated $Li_2Mn_2O_4$. The diffraction pattern of the initial material (**Figure 2 (b)** top) is consistent with $[111]_c$ with the same lattice parameter determined by XRD measurements on the pristine sample material (see Supporting Information). Upon lithiation, the material transforms into the tetragonal structure without rotation of Mn and O octahedra, so that the $<100>_c$ and $<100>_p$ directions are the same. The resultant diffraction pattern (**Figure 2 (b)** top) has fewer spots due to the smaller unit cell[5], and is consistent with the known tetragonal structure. A careful examination of the lamellar structure in the transformed region reveals domains with the three different possible in-plane orientations $<100>_c = <100>_p$, rotated relative to each other by 120°. This is consistent with the three-fold symmetry of the $[111]_c$ zone axis and indicates that the lamellar structures have specific crystallographic orientations.

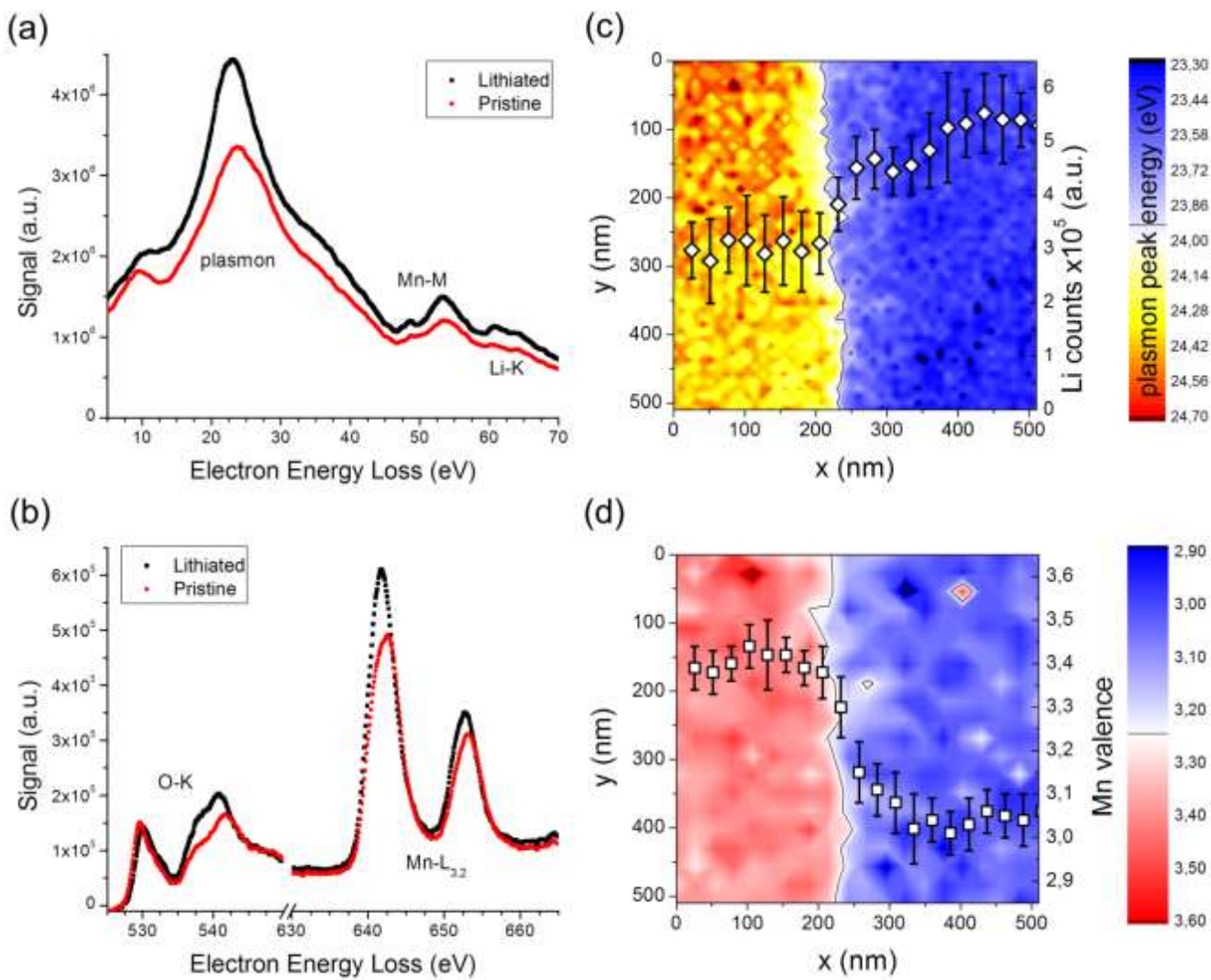

**Figure 3.** EELS of a phase interface region. (a) Dual low-loss EEL and (b) high-loss EEL spectra from the pristine and transformed regions of the sample. (c) Plasmon peak energy map of the phase interface region overlaid with the Li-K edge counts summed along the direction parallel to the interface (error bars are the standard deviation). (d) Mn valence map of the phase interface region overlaid with the Mn valence averaged along the direction parallel to the interface, showing a change from 3.45 to 3.05 across the interface.

A partially in-situ lithiated $Li_xMn_2O_4$ TEM specimen containing a phase interface between $x = 1$ and $x \approx 2$ was characterized using high-resolution TEM and STEM-EELS (**Figure 3**). The electronic structure of the sample is studied in both the low- and high-loss regions of the EEL spectra. The transformation from x = 1 to x ≈ 2 causes the plasmon peak to shift by 0.9 eV to lower energies (**Figure 3 (a)**), which is used in **Figure 3 (c)** to map the interface between the phases. This peak shift is in the same direction but three times larger than the energy shift observed and used to map lithium content in $(Li)FePO_4$[19]. Plasmon peak shifts have also been observed in other transition metal oxides ($Li_{1/2}Ni_{0.5}Mn_{1.5}O_4$[22], various manganese oxides[23]) and indicated a change in electronic properties. An unidentified broad peak present at 10 eV, lying at a similar energy as an intraband transition observed in $(Li)FePO_4$[19], diminishes upon lithiation. Furthermore, the height of the Li-K edge at 58 eV in **Figure 3 (a)** increases by 75%, while the Mn-M edge only slightly changes in intensity while keeping the same shape.

In the high-loss region (**Figure 3 (b)**), the O-K edge, which corresponds to the transition from the O 1s to the hybridized bands of O 2p and Mn 3d[24], shows a shift in the first peak maximum from 529.6 eV to 530.1 eV as a result of lithiation. Additionally, the intensity ratio of the first to second peak of the O-K edge increases. This trend has been observed in multiple EELS studies upon lowering manganese valence due to the interaction of above mentioned states in the $MnO_6$ octahedra.[23,25–29] The Mn-$L_3$ edge moves from 642.7 eV lower to 641.8 eV upon lithiation. Moreover, an increase in the Mn $L_3/L_2$ ratio can be observed also indicating a drop in valence.[23,25–29] The more symmetric shape of the x ≈ 2 Mn $L_3$ peak is consistent with a pure $Mn^{3+}$ state while the shoulder on the x = 1 Mn $L_3$ peak suggests a mixture of $Mn^{3+}/Mn^{4+}$ states. The intensity ratios of the first two oxygen peaks as well as the Mn $L_3/L_2$ intensity ratio have been widely used to map Mn valence changes in the literature. However, intensities are somewhat more sensitive to background signal and detector noise than peak positions.[25] Therefore, the change in absolute energy difference between the O-K and Mn-$L_3$ edges can provide a high signal-to-noise measure of Mn valence.[27]

The position of the phase boundary is clearly shown in **Figure 3 (c)** using a map of the plasmon peak energy. It correlates well with the Li-K peak counts which have been vertically averaged in the map region and overlaid on the map. The lithium counts almost double across the interface, as expected.

The manganese valence map in **Figure 3 (d)** also clearly marks the same position for the interface. The slightly lower resolution follows from higher binning needed to process the data. The change in valence of 0.4(2) from 3.45(12) on the left to 3.05(11) on the right is slightly smaller than the expected 0.5 for the x = 1 to x = 2 transformation, but valence quantification in the TEM and particularly in an in-situ sample is difficult and requires careful calibration standards and/or theory calculations.

The EELS maps of electronic structure at the interface (**Figure 3 (c,d)**) show that the interface is rough, suggesting that the interface is not crystallographically sharp, not planar or not parallel to the electron beam. Thus, the approximate 100 nm interface transition width observed in both lithium and manganese electronic structure and the 50 nm width observed in the plasmon map serve as upper limits of the phase boundary width.

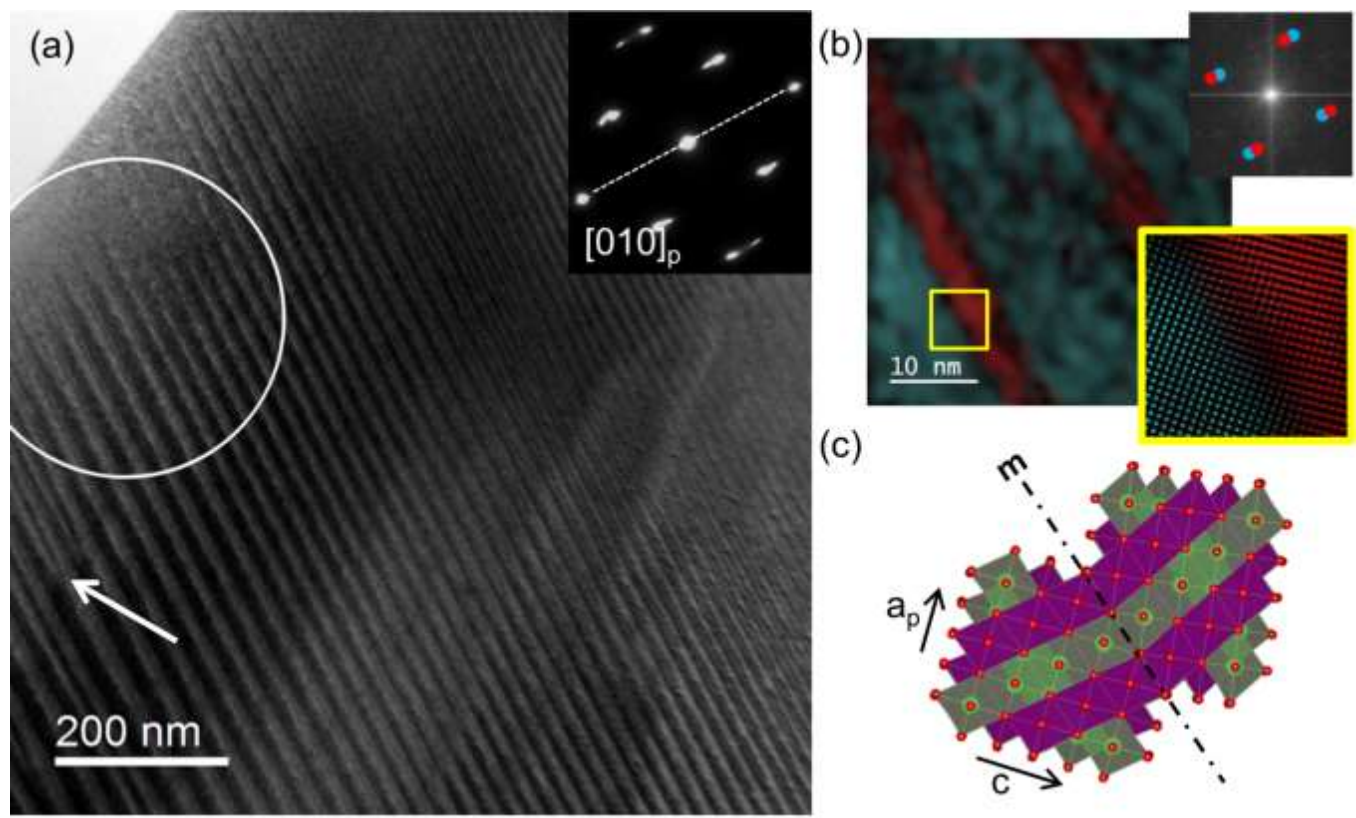

**Figure 4.** Lamellar structure of $Li_2Mn_2O_4$. (a) BF image of an in-situ transformed LMO sample showing lamellar contrast with a periodicity of approximately 30nm. The inset SADP, taken from the circled area at a tilt of ~37°, shows that the contrast stems from two variants of tetragonal $Li_2Mn_2O_4$ that are twinned along a $(112)_t$ plane (indicated by dashed line). (b) Bragg-filtered HRTEM images of the lamellar microstructure in the $[1\text{-}10]_t$ orientation. Twins are revealed by selecting the $\{220\}_t$ (red) and $\{004\}_t$ (blue) planes in the FFT; the twin boundaries are narrow but non-flat. (c) Atomic structure sketch of the $[010]_p$ / $[1\text{-}10]_t$ zone axis showing the orientation and possible position of the twin plane.

Further *ex-situ* analysis of a single domain of the lamellar structure was performed in a double-tilt holder directly after an *in-situ* lithiation (**Figure 4**). The inset SADP (**Figure 4 (a)**) reveals a twinned tetragonal phase with a twinning plane of $(112)_t$ / $(101)_p$, indicated by a dashed line. This is the most common twinning plane in tetragonal minerals. The closely spaced diffraction spot pairs correspond to the $(004)_t$ / $(010)_p$ and $(220)_t$ / $(001)_p$ reflections of the two twin variants. This is confirmed in Bragg-filtered HRTEM images in **Figure 4** (c) that is color-coded using the spot selection shown in the FFT of the image. While the interface plane between both variants is not perfectly planar it is close to the $(101)_p$ twinning plane.

Both BF and Bragg-filtered HRTEM imaging allow the periodicity of the twin variants to be determined as between 20 and 50 nm. The thickness of the twins is not constant across the whole specimen (**Figure 4 (a)** and **Figure 6 (a,b)**); in particular there is evidence for a decrease in twin thickness by twin splitting in regions where the TEM specimen is thinner (see arrows in **Figure 4 (a)**). Note that the lamellar structure seems to completely disappear in the thinnest regions at the edge of the TEM specimen (**Figure 4 (a)**). Scanning nanodiffraction analysis of the twinned sample in **Figure 4** using a beam diameter of 2 nm reveals a majority twin variant thickness of 25-30nm, a minority twin variant thickness of 5-10 nm and a boundary width of 1.5 to 4nm (see Supporting Information).

A sketch of the mirror plane and the two twin variants, excluding any distortions at the interface, provides a possible visualization of the twinned crystallographic structure (**Figure 4 (d)**). It demonstrates that the $(112)_t$/$(101)_p$ twinning plane is parallel to edges of the $MnO_6$ and $LiO_6$ octahedra, allowing them to be joined across the twin boundary in the same way they are joined in the original spinel phase. However, the Jahn-Teller distortion points in different directions in the two variants, resulting in 6° and 9° misalignments between the lattice planes measured in **Figure 5 (b)**. The exact arrangement of the octahedra at the interface and possible deviations in chemical composition could not be detected by HR-TEM, HR-STEM or STEM-EELS in our *in-situ* samples.

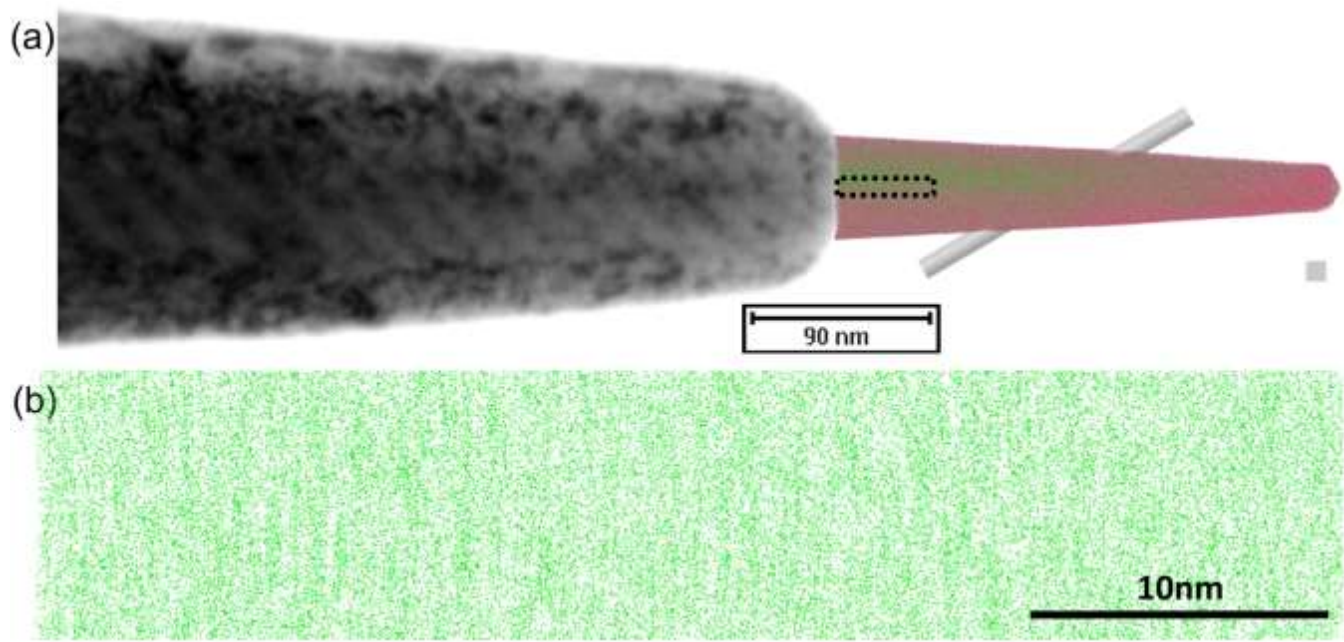

**Figure 5.** APT analysis of *in-situ* lithiated $Li_2Mn_2O_4$. (a) BF-TEM image of the needle after APT analysis of the specimen from **Figure 4** and the APT reconstructed volume (Li: green, Mn: red). Analysis of the local lithium concentration along the 10 nm diameter cylinder perpendicular to the twin planes showed no evidence of lithium heterogeneity (Supporting Information). (10 nm box for scale) (b) Planar slice of a part of the reconstructed volume closest to the lamellar contrast still visible at the tip (rectangle in (a)) showing a homogeneous lithium distribution.

Laser-assisted atom probe tomography (APT) is used to look more carefully for any possible variations in composition within the twinned structure. We use here the same APT method that has been successfully applied previously to investigate lithium distributions in LMO by Pfeiffer and Maier, et al.[30,31] In this case, a fine needle was cut directly from the *in-situ* TEM lithiated specimen in **Figure 4** using FIB, and then analysed with laser-assisted APT (**Figure 5**). The Li to Mn ratio is roughly double of what has been found in previous APT reconstructions of $LiMn_2O_4$, further confirming that the *in-situ* TEM lithiated material is the $Li_2Mn_2O_4$ phase. The reconstruction of the APT analysis of the twinned $Li_2Mn_2O_4$ specimen showed no evidence of chemical heterogeneity (**Figure 5 (a)**). In particular, a careful analysis of the lithium composition averaged perpendicular to the twin planes (along the cylinder sketched in **Figure 5 (a)**) showed no evidence of heterogeneities above the noise level (see Supporting Information). This is additionally visualized in **Figure 5 (b)** showing a homogeneous lithium distribution in a planar slice of the reconstructed volume directly at the tip of the remaining needle (dotted rectangle), where lamellar contrast is still visible in BF-TEM image (**Figure 5 (a)**).

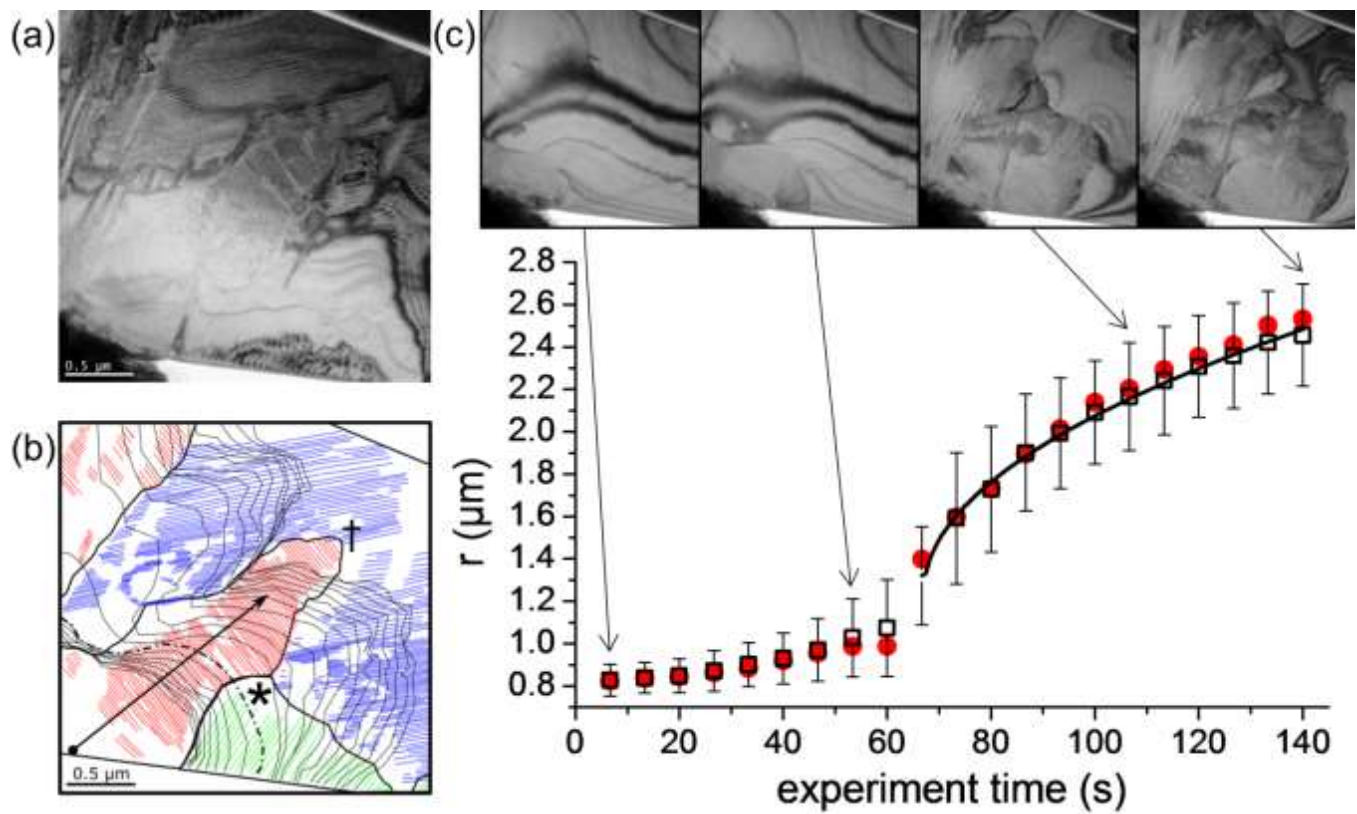

**Figure 6.** Lithiation kinetics. (a) A bright field image of the specimen from Figure 2 after lithiation is complete. (b) Position of the reaction front for every 20th frame (ca. 7 s) of the image series (thin black solid lines). The twin boundaries are marked where they could be detected in different colors to indicate the different domains. Domain boundaries are marked with thick black solid lines. The contact point with the lithium tip is indicated by a dot and the position of the front at 60 s is indicated by the dashed-dotted line. (c) The evolution of the mean

(black squares) and median (red circles) distance of the reaction front from the contact point. The solid line shows a square root fit to the data for times greater than 60 s.

We have investigated the kinetics of the reaction by recording the rate of propagation of the phase interface away from the contact to the Li tip. **Figure 6** provides an overview of the reaction front motion for the *in-situ* lithiated specimen shown in **Figure 2 (a)**. A TEM image of the fully lithiated specimen is shown in **Figure 6 (a)**, while a sketch showing the progression of the phase interface, the resultant domain structure and the twin spacing and orientations are shown in **Figure 6 (b)**. One sees that the reaction front velocity varies strongly from position to position along the boundary, leading to a rough interface. Nonetheless, a plot of the mean and median distances of the reaction front from the contact point shows a clear trend with time (**Figure 6 (c)**). Due to the fact that the video was only started after contact between the lithium tip and the sample had already been established, the reaction front had already moved around 0.8 µm into the sample. After this, the bias was ramped up at 0.1 V/s to reach -5 V within 60s and the boundary arrived at the position shown by a dotted-dashed line in **Figure 6 (b)**. After a stable bias of -5 V is reached, the boundary moves more rapidly but with a steadily decreasing velocity. This behavior is well fit with a square root function (**Figure 6 (c)**), indicating that the reaction front movement is diffusion limited. The square root fit also yields an estimate for the diffusion coefficient of lithium in the transformed $Li_2Mn_2O_4$ region. Setting the boundary position equal to $r(t) = \sqrt{D(t - t_0)} + r_0$ using a linear approximation, we obtain an estimate of $D = 2 \cdot 10^{-10}$ cm$^2$/s. This is in reasonable agreement with reported values for bulk diffusion of lithium in polycrystalline $LiMn_2O_4$, $D = (3.4 \cdot 10^{-8} - 1.71 \cdot 10^{-12})$ cm$^2$/s, where single crystalline samples tended to have smaller diffusivities.[32] This is also multiple orders of magnitude smaller than the surface diffusion coefficient of lithium on graphite ($D = 5 \cdot$ $10^{-6}$ cm$^2$/s)[33], which allows us to conclude that the reaction front is controlled by lithium diffusion in the twinned single crystal Li$_2$Mn$_2$O$_4$ phase.

We analyzed the data in **Figure 6 (b)** in detail to understand the reasons for the strongly heterogeneous interface velocities. The propagation of the reaction front is strongly hindered at the interface between the red and upper blue domains, but not at any of the other interfaces, so that a connection between interfaces and front propagation is not immediately obvious. Plots of the local interface velocity (magnitude and direction) as a function of local twin orientation also did not show a clear correlation (Supplementary Information). We also found no systematic effect of the estimated local foil thickness (which is thinner near the edges) or experiment time on either the local front velocity or the twin spacings. In contrast, *in-situ* lithiation of several other samples showed that existing stacking faults in the LiMn$_2$O$_4$ strongly hindered front propagation. The specimen shown in **Figure 6** did not contain stacking faults in the pristine state, however the presence of dislocations and point defect clusters cannot be ruled out. Thus, we presume that the main factor causing the locally variable front propagation velocities are the presence of defects and stresses that we could not identify during pre-characterization.

In total, 15 *in-situ* TEM lithiation experiments were performed. All showed a sharp phase interface that propagated into the TEM specimen and the formation of a twinned structure in the transformed region. Between one and 3 different oriented twinned domains were observed with twin thickness periodicities between a few and 100 nm. A similar twinned microstructure was found in the tetragonal Li$_2$Mn$_2$O$_4$ phase contained within several of the investigated pristine LiMn$_2$O$_4$ particles (see Supporting Information). This indicates that the twinned microstructure can occur in lithium manganese spinel even without the extreme conditions used for the *in-situ* experiments (large overpotential, high C-rate and TEM irradiation). In particular, it shows that the minimum specimen dimensions (ca. 100 nm thick TEM foil versus 5 μm diameter particle) do not completely control phase stability, in contrast to the interpretation of electrochemical measurements in a previous report.[16] Generally, twinned microstructures with very similar microstructures result from stress-driven, diffusionless martensitic transformations from the cubic to tetragonal phase in a wide range of materials with similar c/a ratios. This introduces the idea that the martensitic-like microstructure observed here may be the result of stresses created by lithiation. A similar twinned microstructure in lithium containing manganese oxide has to our knowledge only been reported in oxygen-deficient $Li_{1+\alpha}Mn_{2-\alpha}O_{4-\delta}$, which showed tetragonal distortion directly after calcination,[6] while twinning is not uncommon in other types of manganite spinels.[34,35]

The TEM foils have a roughly equi-biaxial plane stress state, so that stress-driven twinning should show a clear dependence on the out-of-plane orientation of the sample. The simplest test of this idea, which neglects stresses due to defects and accommodation stresses at the twin interfaces, is to compare the out-of-plane crystallographic orientation of the foils with the number of domains. If stress relief due to the plane-stress state of the foil is the dominant effect, then the c-axis should point as far as possible out of the plane, and the foils with near $<111>_c$ orientations, having all possible c-axes equally pointed out of plane, should contain three domains (e.g. **Figure 2**), while foils with orientations further away from $<111>_c$ should contain one (e.g. **Figure 4**) or two domains. A comparison of all 15 *in-situ* samples showed some tendency for the twin domains to form as predicted by stress relief (see Supporting Information). However, evidence that driving forces other than the foil plane stress state are important can be seen directly in **Figure 6 (b)**. During reaction of the initially single crystal region, a new domain is formed (indicated by * in **Figure 6 (b)**), and another domain ceases to grow (indicated by †).

Since the initial plane stress state due to the foil geometry is homogeneous, we presume that the reasons for the observed domain selection during growth must be due to the interaction of the interface stresses[36] with defects that could not be seen during pre-characterization.

We found no clear connection between the twin spacings and the out-of-plane foil orientations. For example, if the foil plane stress state is decisive in driving twin formation, then twins are less likely to form in foils with an out-of-plane orientation along one of the <100>$_c$ cubic directions, since the 13% expansion due to the lithium induced cubic-to-tetragonal transformation can occur unconstrained, leaving only the ca. 1% contraction to be accommodated by in-plane stresses. However, we unfortunately never encountered a near <100>$_c$ out-of-plane oriented sample, and it is possible that strains on the order 1% are already more than sufficient to initiate twinning as a stress relief mechanism.

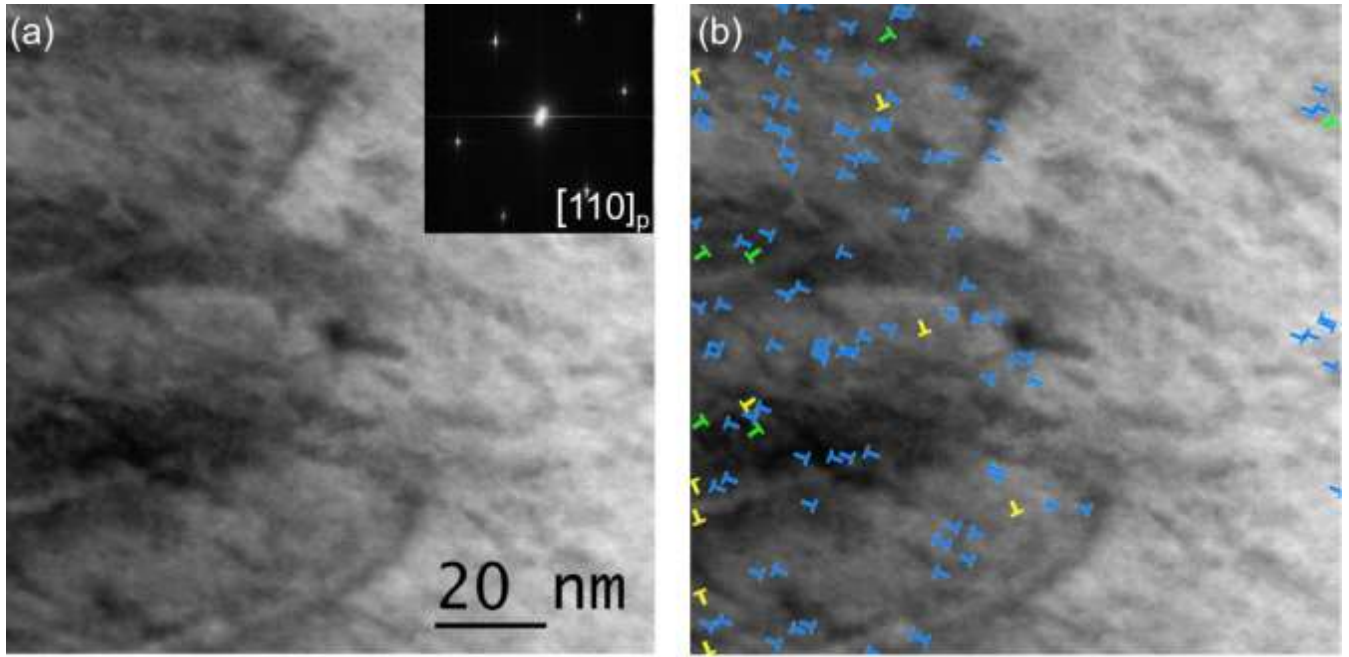

**Figure 7.** Microstructure at the phase interface. (a) A high-resolution BF-STEM image in $[1\bar{1}0]_p$ orientation (see FFT inset) reveals lenticular-shaped needles at the phase boundary, typical of mechanical twinning. (b) Single plane Bragg filtering allows three different dislocation types to be identified, indicated by three different colors including the direction of the inserted half-plane. The dislocations are mostly localized in the left, transformed region of the sample.

**Figure 7** shows a high-resolution STEM image of the phase boundary. The serrated structure of the interface reveals the lenticular-shaped needle structure of the transformed material, which is typical for mechanical twinning. The twins themselves are not discernible by Bragg filtering since the twinning plane is not parallel to the beam as a consequence of the $[110]_p$ orientation of the sample. FFT analysis of the image shows local variations in the c/a-ratio and an average value of 1.12 which is 25% smaller than the expected value of 1.16. Similar local deviations from the expected c/a-ratio have been observed in an Al-doped LMO and attributed to ferroelasticity and defects.[37] The sharp localization of bending contours at the interface plus blurring of the FFT spots provide evidence for strains near the interface. Nevertheless, Bragg-filtering of the image along the three principal directions can be used, see **Figure 7 (b)**, to examine the coherency of the boundary by looking for interrupted planes. This shows that the left, transformed region is heavily populated by dislocations of mostly one Burgers vector direction, while only a few dislocations are located further ahead of the boundary in the untransformed material. This allows us to conclude, that although twinning presumably relieves some of the transformation stresses, it still involves dislocation formation and generation of a semi-coherent interphase interface.

The observation of both Li diffusion controlled motion of the interface and of the martensitic needle structure characteristic of mechanical twinning suggests that the transformation is a coupled diffusional/displacive transformation.[38] The differences to purely displacive transformations like the martensitic transformation are therefore in the kinetics due to the diffusional component. But the crystallographic martensitic transformation theory is still applicable, which also predicts dislocations in the planar interface between cubic and tetragonal phase.[39] Displacive transformations are often reversible, with the most notable case being shape memory alloys, where the boundary between both phases is highly glissile. However, a significant loss in reversibility is observed once the transformation causes plastic deformation and dislocation accumulation.[40,41] The highest reversibility is reached when the crystallography allows an undistorted habit plane between both phases and when there is no volume change to the unit cell[42]. For steel, the large volume change during the martensitic transformation is often given as the reason for irreversibility.[43]

In light of the large volume change $\Delta V \approx 5\text{-}6\%$ associated with lithiation of LMO and the observed dislocation accumulation near the phase boundary (**Figure 7**) a highly reversible transformation is not expected. In fact, attempts to reverse the movement of the lithiation front were not successful even at very high overpotentials up to +15V. It is conceivable that the applied overpotential is not large enough to drive the *in-situ* reaction, as a result of charging from secondary electron emission[20] and possible potential barriers to moving the Li through the lithium oxide/nitride layers[44]. Nonetheless, successful delithiation of α-MnO$_2$ nanowires has been reported using a similar *in-situ* setup with a solid lithium electrode at a potential of +4V[45]., suggesting that the origin of the irreversibility observed here lies in the transformation itself and not in experimental complications.

**Summary and conclusions**

We have demonstrated that *in-situ* lithiation of Li$_x$Mn$_2$O$_4$ in the TEM leads to the expected phase transformation from cubic spinel (x = 1) to tetragonal spinel (x ≈ 2) and that the two phases are clearly distinguishable by a sharp semi-coherent interface. We were able to verify the successful lithiation by both structural and spectroscopic analysis of the lithium content and manganese valence, but did not detect the orthorhombic intermediate phase previously reported in nanowires.[15] The deviations from the expected stochiometric values for the lithium content ratio (1.75 instead of 2) may be explained by the *in-situ* conditions and the difficulty of detecting Li, but is also consistent with a valence change of only 0.4. Furthermore, it is possible that Li vacancies are stable in the tetragonal phase, extending the equilibrium composition range to below x = 2, in which case the measured composition at the reaction front would be less than x = 2, reflecting a finite solubility of the x = 1 phase in the x = 2 phase.[4,16,46] The small measured deviations from the expected manganese valences are well within the typical range of accuracy for EELS, and depend critically on the method used to determine valence from the electron energy loss spectra. In fact, the possibility here to perform manganese valence determination across a chemical boundary within a single sample offers a rigorous test of the various methods to determine valence; in LMO the energy difference between the O-$K_a$ edge and Mn-$L_3$ edge seems to be more reliable, rather than the more widely used methods based on the Mn-L edge intensities. By analyzing reaction front movement, we conclude that the reaction is diffusion limited and were able to obtain a value for the lithium diffusivity in twinned single crystal $Li_2Mn_2O_4$, which is in the range of previously reported values for $LiMn_2O_4$, indicating that transport in the tetragonal phase is not a rate-limiting factor during battery operation. Our measurement provides a reference value of the diffusivity that can be used to identify roles of grain boundaries, particle ensembles, polycrystallinity and cycling damage that have influenced previous measurements of rate-limiting transport.

During the *in-situ* tetragonal transformation of the lithium manganese oxide, we observed the formation of a complex twinned microstructure with a well-defined crystal orientation relative to the initial cubic phase and strongly reminiscent of mechanically twinned microstructures. We conclude that the twin microstructure formation is stress-driven, although the lack of a clear dependence on the plane stress state of the foil not made, points to the importance of stress fields from defects and from phase misfit. This twinned tetragonal phase could not be reversibly transformed back to the cubic phase, presumably due to the accumulation of dislocations at the phase interface. Residual tetragonal domains have been previously found in the TEM in 0<x<1 cycled samples and attributed to local overpotentials and named as a cause for battery fatigue[11]. Therefore, we suggest that the twinned microstructure hinders delithiation of the tetragonal phase and contributes to capacity loss in $Li_xMn_2O_4$-based batteries.

**Methods**

**Sample preparation.** Sigma Aldrich lithium manganese (III,IV) oxide $LiMn_2O_4$ powder with a nominal particle size smaller than 5µm was used. Large, facetted particles were selected and cut into TEM foils using a Focused Ion Beam microscope (FIB, FEI Nova Nano Lab 600 FIB-SEM). After depositing a Pt protective layer, a micron-sized block aligned to the main particle facets was transferred to a TEM half ring (Omniprobe grid) using a micromanipulator. There, one corner was thinned to TEM transparency with the FIB. The Pt protective layer was later removed from the edge of the foil to make the specimen edge accessible to contact by the lithium tip.

**TEM characterization/*in-situ* method.** Pre-characterization of the specimen was performed using a 300 keV TEM (Phillips CM30) to confirm the $LiMn_2O_4$ cubic spinel structure and characterize any grain boundaries or other defects. *In situ* experiments and subsequent *ex situ* characterization were conducted in a FEI Titan microscope at 300kV with an image corrector and a Gatan GIF Quantum with Dual EELS capability.

The *in-situ* holder is a piezo-controlled Nanofactory STM-TEM single-tilt holder in which the FIB-prepared specimens were mounted. Immediately before inserting the holder into the TEM, the W tip of the STM-holder was scratched across a freshly cut Li surface covered in liquid n-pentane (ChemSolute). The n-pentane slows oxidation of the lithium while the W tip is mounted in the holder, the holder is inserted into the TEM, and pumped down, all in less than a minute. The n-pentane keeps the lithium from completely reacting, but allows a layer of lithium oxide/lithium nitride to form on the surface, which acts as a solid electrolyte during the experiment.[47,48] Lithiation inside the TEM is initiated by mechanical contact between the tip and the specimen; the contact is confirmed by movement of the specimen, bending contours observed in bright field mode and small changes in the electrical currents. A bias of -5V is applied between tip and sample to drive the reaction and counteract possible positive biases in the sample due to the secondary electron generation and the lithium oxide barrier.[20,21] The reaction could be halted by breaking off the contact between sample and tip.

**EELS characterization of the specimen.** Energy drift correction and deconvolution of the low-loss EEL spectra were performed using built-in functionalities of Digital Micrograph. Li content near the reaction interface was obtained by mapping the lithium K-edge, integrated over a window of 15eV behind the edge after a power-law background subtraction had been performed. Since all acquisition parameters were the same within a given map, the edge counts are proportional to the number of atoms in the beam path and therefore provide a direct measure of relative changes in lithium content[49], assuming that other element compositions do not vary during an experiment. The plasmon position was determined by a Gaussian fit to a 5eV region surrounding the maximum.

Manganese valence was determined using a method established by Zhang et al. for manganese oxides based on the energy difference between the O-$K_a$ edge and Mn-$L_3$ edge.[27] The energy difference was measured after of power-law background subtraction based on the background in front of the O-K edge and deconvolution using Digital Micrograph. We note that this method gave valences closer to the expected values than methods based on the Mn-L edge intensities. The error for this method is taken as the approximate 3.5% percent of the value stated in the above publication, since the standard deviations calculated from our data are about equally large.

**Atom probe tomography analysis and sample preparation.** A tip with ca. 30 nm apex radius was prepared by FIB from a previously *in-situ* lithiated TEM specimen that was interim stored under vacuum. After transfer of the sample to a tungsten support wire, the tip was cut and sharpened using an established lateral cut method.[50] Laser-assisted atom probe analysis was performed using a method developed for $LiMn_2O_4$.[31] The analysis temperature of 30 K immobilizes the lithium and allows determination of lithium concentrations at and across defects.[30]

ASSOCIATED CONTENT

Video of *in-situ* experiment referred to in **Figure 1 (a)** and **Figure 6** (AVI);

Supporting Information document referred to in the article (PDF)

AUTHOR INFORMATION

**Corresponding Author**

* Institute of Materials Physics, University of Göttingen, Friedrich-Hund-Platz 1, 37077 Göttingen, Germany

Tel. +49(0)-551 39-5011; volkert@ump.gwdg.de

**Author Contributions**

T. Erichsen performed the TEM experiments including the sample preparation and analyzed the associated data and did most of the writing. B. Pfeiffer performed the APT-sample preparation and characterized as well as analyzed the according data. V. Roddatis suggested important improvements to *in-situ* experiments, supported post-mortem sample investigations and contributed to data analysis. C. A. Volkert conceived the project and wrote part of the manuscript. All authors discussed the results.

**Funding Sources**

Deutsche Forschungsgemeinschaft [CRC 1073 - Project C05 and Z02]

ACKNOWLEDGMENT

The crystal structures in this work were sketched with VESTA.[51] Help in the display of the APT data by J. Arlt is gratefully acknowledged; we also thank C. Nowak for developing the APT methods used for analyzing LMO. We would also like to thank C. Borchers and Qu R.-T. for their critical review and helpful discussion. This work was funded by the Deutsche Forschungsgemeinschaft [CRC 1073 - Project C05 and Z02]. The use of equipment in the "Collaborative Laboratory and User Facility for Electron Microscopy" (CLUE) www.clue.physik.uni-goettingen.de is gratefully acknowledged.

# Supplementary information for "Tracking the diffusion-controlled lithiation reaction of LiMn$_2$O$_4$ by in-situ TEM"

*Torben Erichsen, Björn Pfeiffer, Vladimir Roddatis, and Cynthia A. Volkert*

The video file that is the basis for the still image in **Figure 1 (a)** and the series of still images in **Figure 6** is separately uploaded. The video was recorded over a time of 330 s and is processed to play at 40x speed at 12fps with a field of view of 2.7 µm. To fulfill the size constraints of the publisher the video only contains every 10$^{th}$ frame of the original in-situ data and has additionally been binned by two in both vertical and horizontal direction. To assess the quality of the original data single frames are referenced within above mentioned figures. The video was only started after contact between the Li tip and the sample had been established and the reaction front had already progressed about one micron into the sample.

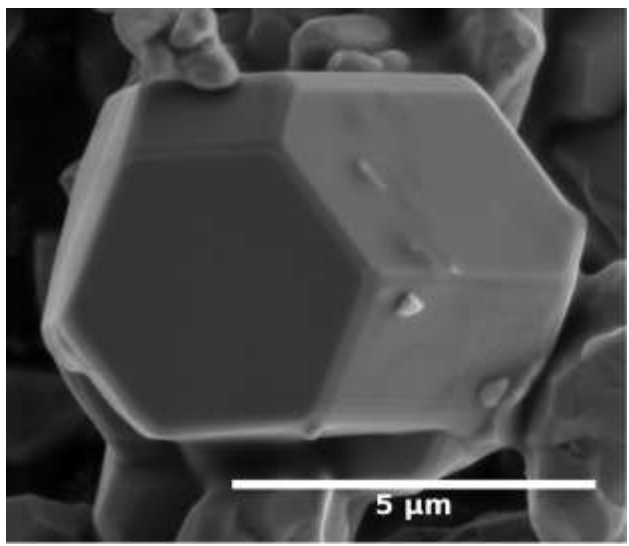

**SI-Figure 1.** SEM image of a typical particle prior to sample preparation. Large particles of 4 µm or larger with well-defined facets were selected for lamella preparation.

The pristine sample material as obtained from Sigma Aldrich has a wide particle size distribution with the smallest particles being on the order of hundreds of nanometers and the largest ranging up to multiple microns. The large particles have clearly faceted shapes: some are highly symmetric (e.g. **SI-Figure 1**) suggesting they are single crystals while others appear to be composed of a few crystallites. The single crystal particles were chosen for sample preparation and the lamella aligned along the longest axis to ensure a large TEM specimen but also to improve the odds of obtaining an orientation close to a zone axis, since the single-tilt TEM holder used for the in-situ lithiation experiments does not allow easy specimen alignment.

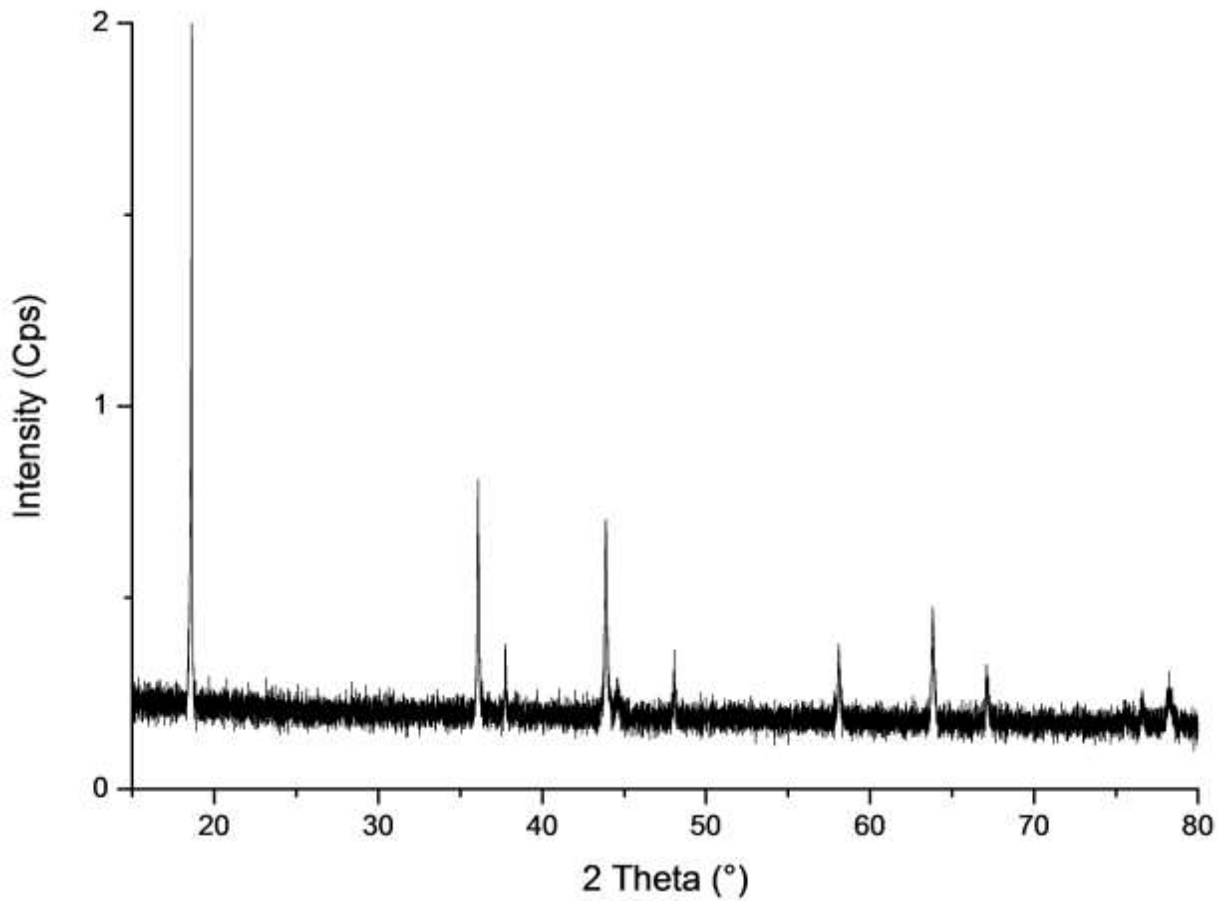

**SI-Figure 2.** θ-2θ XRD data from the $LiMn_2O_4$ particles. The peaks fit the expected cubic spinel structure with a measured lattice parameter of a = 8.247(4) Å.

The XRD characterization presented in **SI-Figure 2** shows the expected peaks for cubic spinel $LiMn_2O_4$ with a lattice parameter a = 8.247(4) Å, in agreement with reported literature values. Additional peaks can clearly be assigned to the aluminum SEM stub below the powder.

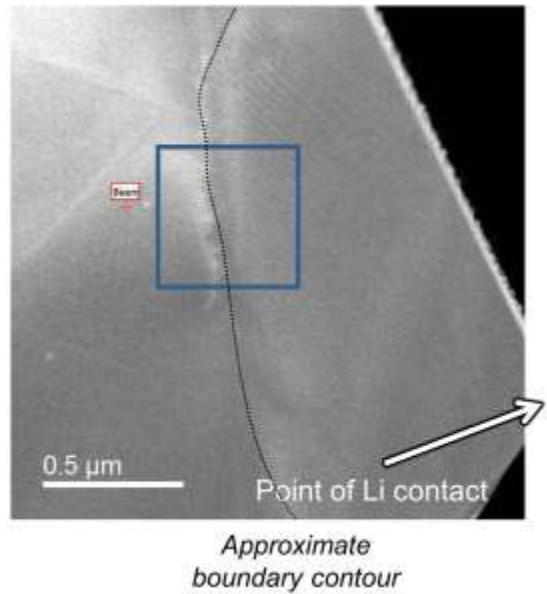

**SI-Figure 3.** Overview of the sample used for the EELS mapping at a stopped reaction front. Lamellar contrast is visible in the tetragonal phase on the right.

   **SI-Figure 3** shows the TEM lamella geometry and gives an overview of the region used for the detailed EELS data analysis presented in the paper. The approximate position of the front is marked (dotted line) to show that it is not perfectly planar.

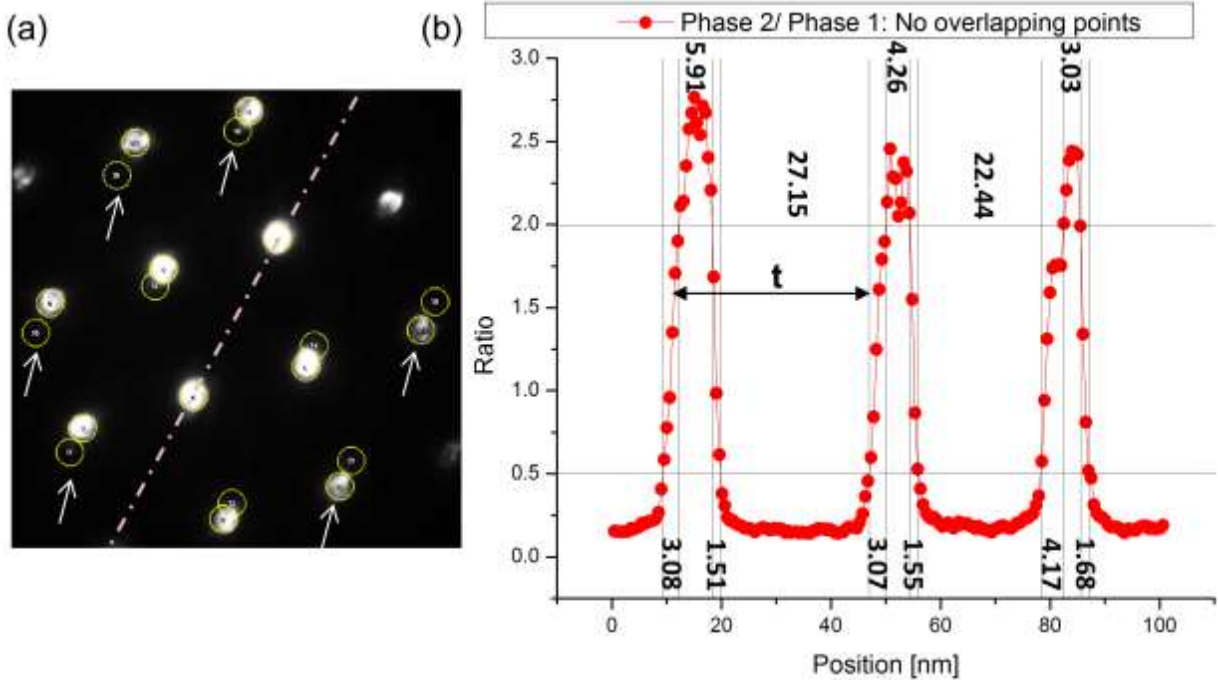

**SI-Figure 4.** Nano-diffraction of the lamellar structure. (a) A typical nano-diffraction pattern reveals a twinned microstructure with a $(101)_p/(112)_t$ twinning plane (dash-dotted line). (b) A nano-diffraction line scan allows the widths of the twin variants and the twin boundaries to be determined. The thickness of both twins, i.e. twin spacing **t** is indicated by an arrow.

Scanning nano-diffraction with a 2 nm diameter beam and large convergence angle (to protect the CCD) was performed on a $[010]_p$ aligned sample with a large single twinned domain, **SI-Figure 4**. Twin identification was performed using the ratio of the integrated intensities of the diffraction spots (yellow circles) from each twin variant; only spots were used that included intensity from one twin variant (arrows). A plot of the ratio of the integrated intensities along a linescan perpendicular to the twin planes allows the twin variant widths to be determined. The thicker variant has a width between 22-28 nm; the thinner variant has a width between 3-6 nm. The twin boundary widths alternate between 1.5 nm and 3-4.5 nm. This indicates that the 1.5 nm thick boundaries are parallel to the roughly 2 nm diameter electron beam and are planar over this length scale. In contrast, every second twin boundary has a measured width that is larger than the electron beam diameter, indicating that these boundaries are not perfectly parallel to the electron beam or are have 3-4.5 nm roughness.

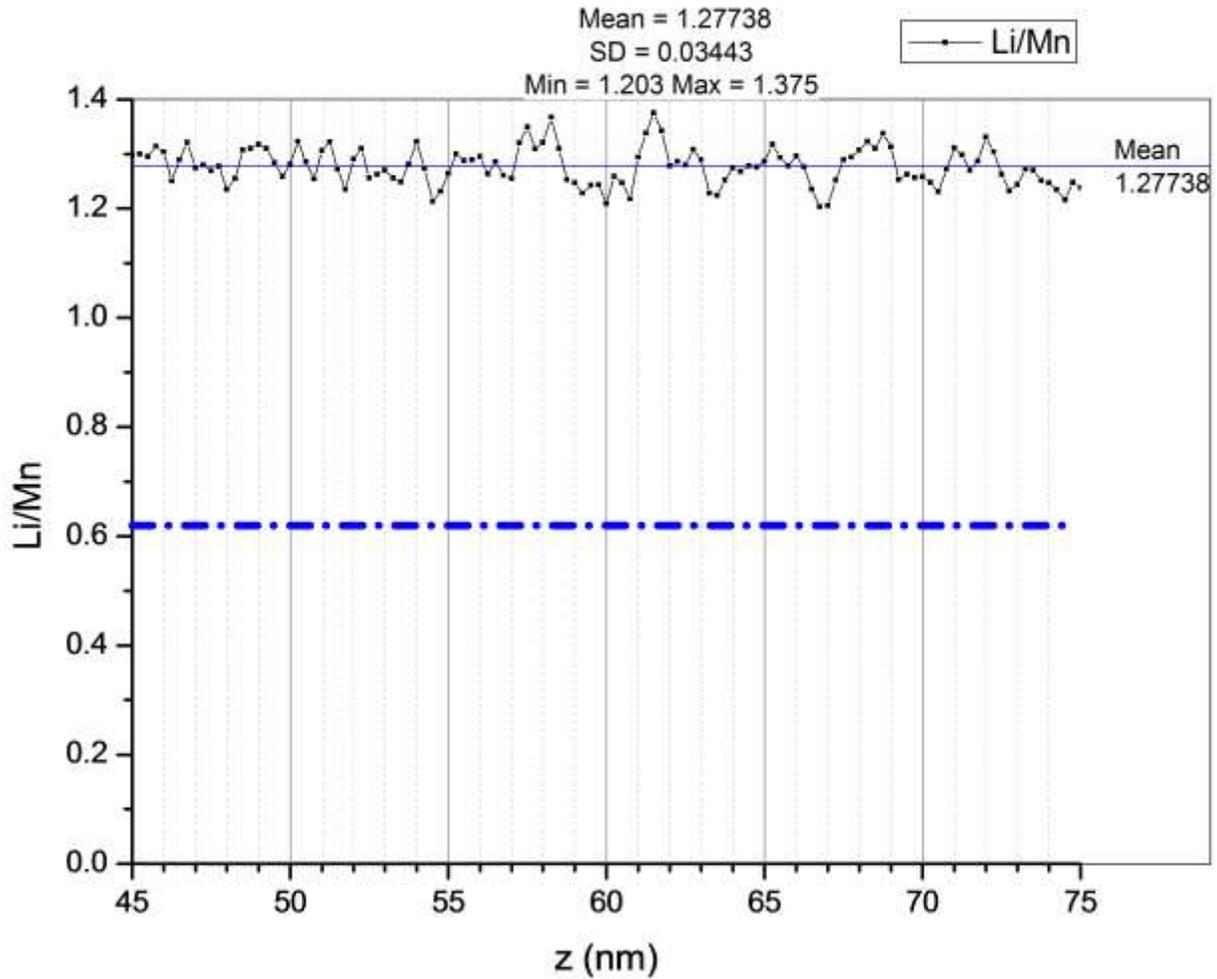

**SI-Figure 5.** Local lithium concentration analysis by APT. The Li/Mn ratio measured in the transformed tetragonal material is approximately twice the value measured in the pristine $LiMn_2O_4$ samples (dash-dotted line).

**SI-Figure 5** shows the local Li/Mn ratio obtained from an APT reconstruction (**Figure 5**) averaged along a cylindrical volume that is aligned perpendicular to the twin boundaries. The mean ratio of 1.28 is twice as large as the mean ratio measured in the pristine starting material 0.615.[31] This was however measured in a high quality conventional APT sample with a presumably much larger number of counts (~10 million), therefore having a much smaller standard deviation. APT analysis does not generally yield absolute compositions due to lost counts and the formation of complex ionized molecules; however, ratios of element counts are considered to be a reliable and quantitative measure of changes in composition by Maier, et al.[31] The local variations in the Li/Mn ratio in the twinned material doesn't exceed 8% from the average and doesn't correlate with twin boundary positions or spacing. Thus we conclude that with our resolution of 1 nm the depletion or segregation at the twin boundaries doesn't exceed this value. Based on the standard deviation there is no indication that the two of the $Li_2Mn_2O_4$ differ by more than 3%.

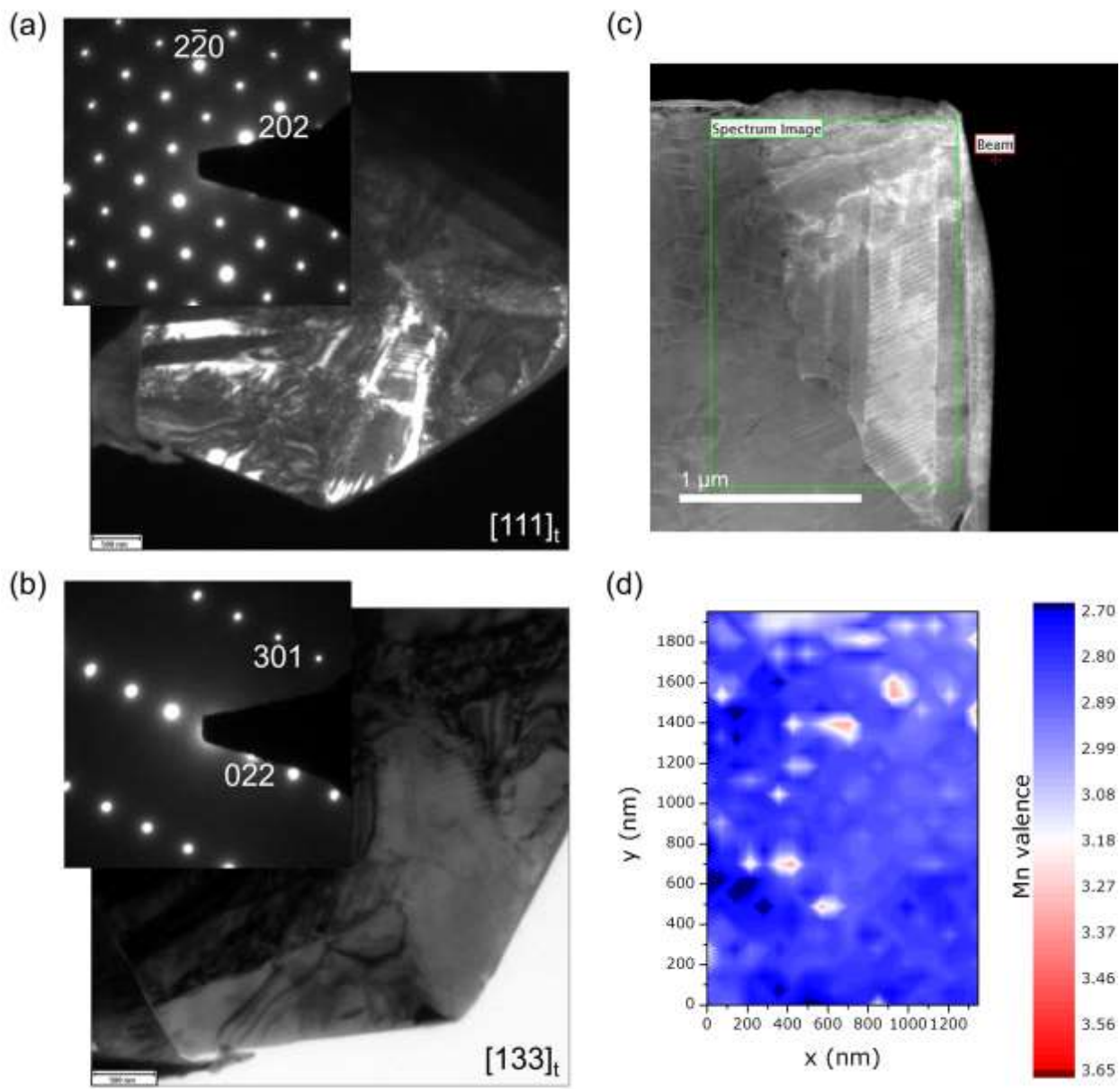

**SI-Figure 6.** Pristine particles containing the twinned tetragonal $Li_2Mn_2O_4$ phase. (a,b) Dark and bright field images including associated diffraction patterns from a pristine particle containing the tetragonal phase with a lamellar microstructure that is comparable to the in-situ transformed material. (c) Bright field image of a second pristine particle containing a volume with lamellar microstructure. (d) STEM-EELS mapping from the region shown in (c) gives a value of the manganese valence close to 3.0 which is consistent with the tetragonal phase.

Despite the fact that the XRD data in **SI-Figure 2** did not show any signs of an additional phase in the pristine LiMn$_2$O$_4$ powder, several particles containing tetragonal lithium manganese oxide were found during the TEM investigations. **SI-Figure 6 (a,b)** shows tetragonal diffraction patterns from one of the particles. The images in **SI-Figure 6 (a,b,c)** all show a clear lamellar contrast similar to what is observed in the in-situ transformed material. STEM-EELS mapping **SI-Figure 6 (c,d)** confirms that the entire mapped region including lamellar structure have a Mn valence much closer to 3.0 than 3.45, the valence measured in the unlithiated region of the in-situ sample presented in the paper. This leads us to conclude that the lamellar nanotwinned structure is a common microstructure for tetragonal Li$_2$Mn$_2$O$_4$, presumably formed by stress relief, and not just a result of the extreme conditions found in the TEM.

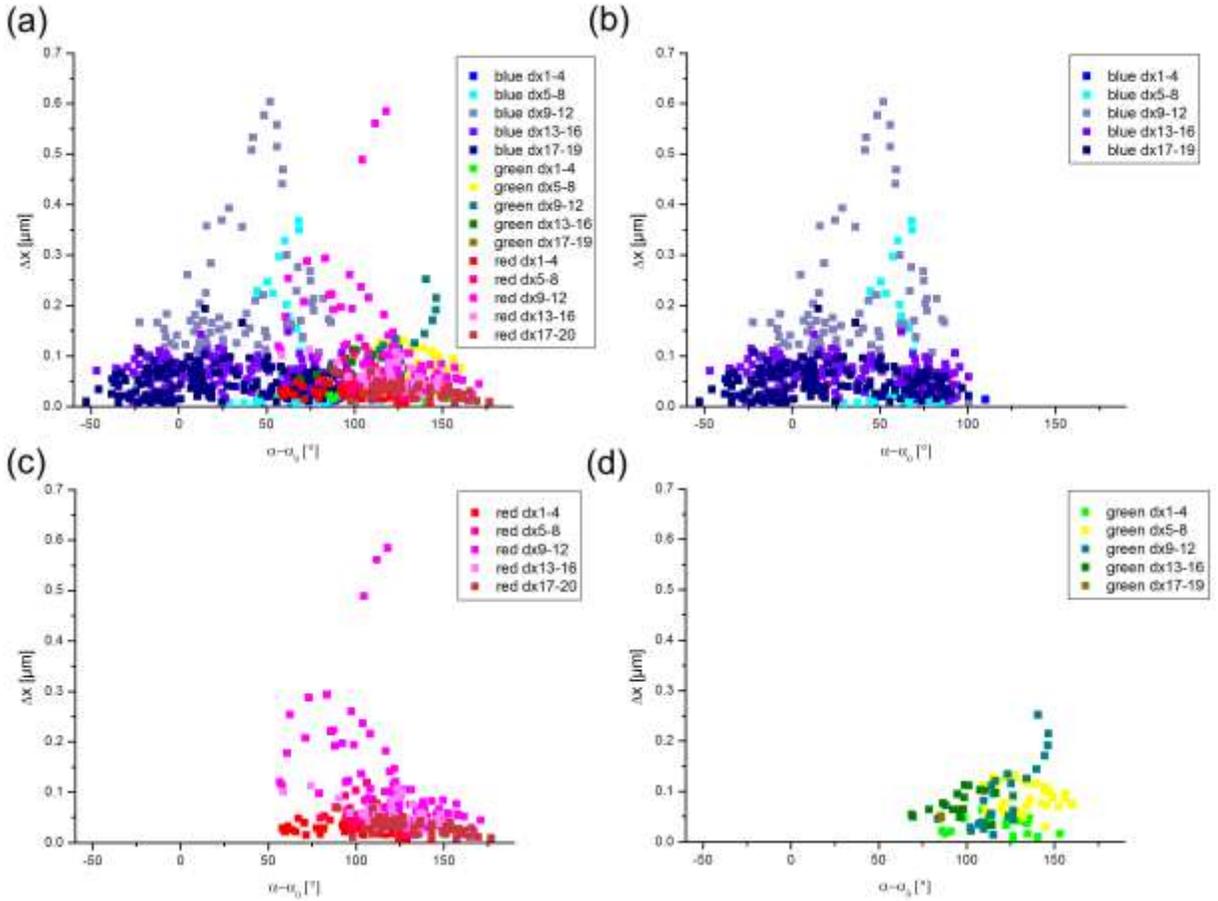

**SI-Figure 7.** Directional dependence of front movement. The Δx of the reaction front for the fixed time steps of the in-situ experiments has been mapped for four intervals. The x-axis corresponds to the difference between the local angle of the reaction front movement α and the direction of the twin boundary normal $α_0$. (a) shows all data as well as the data of the blue domain (b), the red domains (c) and the green domain (d) separately.

As discussed in the paper we did not find a clear correlation between the orientation of the twin boundaries and the direction of fastest lithiation front movement. The data that was collected for this assessment is presented in **SI-Figure 7**. The displacement of the boundary during 4 different time intervals is plotted for the different domains as a function of the angle between the direction of reaction front movement and the twin boundary normal. Despite all three domains exhibiting a direction of fastest motion, which roughly corresponds with the average diffusion direction for the domain, there is no correlation with the twin boundary orientations. Competition between domains as well as the fact that the propagation directions are limited to a range of only 90° due to the sample geometry, likely limit how rigorously we can investigate possible correlations between the tetragonal phase microstructure and the front velocity. In any case, the lack of a clear dependency suggests that possible fast diffusion paths for Li, such as along twin boundaries, are not dominant.

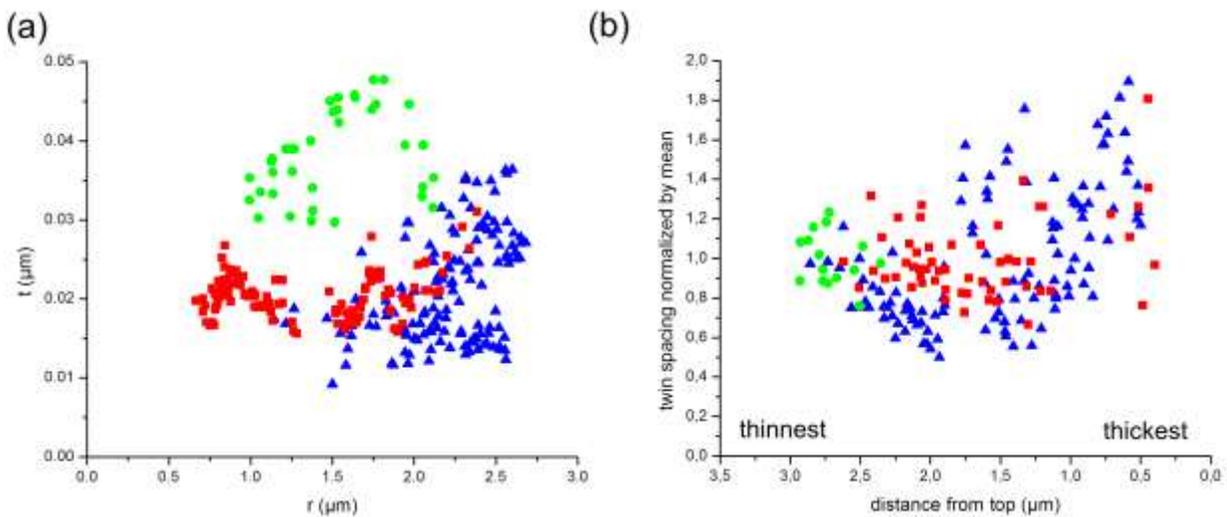

**SI-Figure 8.** Different scatter plots to evaluate the influence of various parameters twin spacing for the specimen in **Figure 2** of the paper. The (normalized) twin variant thickness is plotted against the distance r to the lithium contact point **(a)**, which depends on experiment time, and against the distance from the top of the lamella image (presumed to scale linearly with the lamella thickness) **(b)**.

We further investigated the influence of different parameters on the variation in twin spacing between different domains and within single domains. **SI-Figure 8** shows scatter plots aimed at looking for correlations between the twin spacing and the diffusion distance **(a)** and the TEM foil thickness (expected to increase roughly linearly with distance from the free edge of the foil as a result of the FIB preparation method) **(b)**. Normalization has been performed by the average. The data of the green domain shows no trends for either of the parameters and plots possibly because it was confined to a small region of the sample. However, even though one might see a slight trend in both the blue and red domain the data is too noisy to come to any conclusions not even whether the paramaters in **(a)** or **(b)** are more relevant.

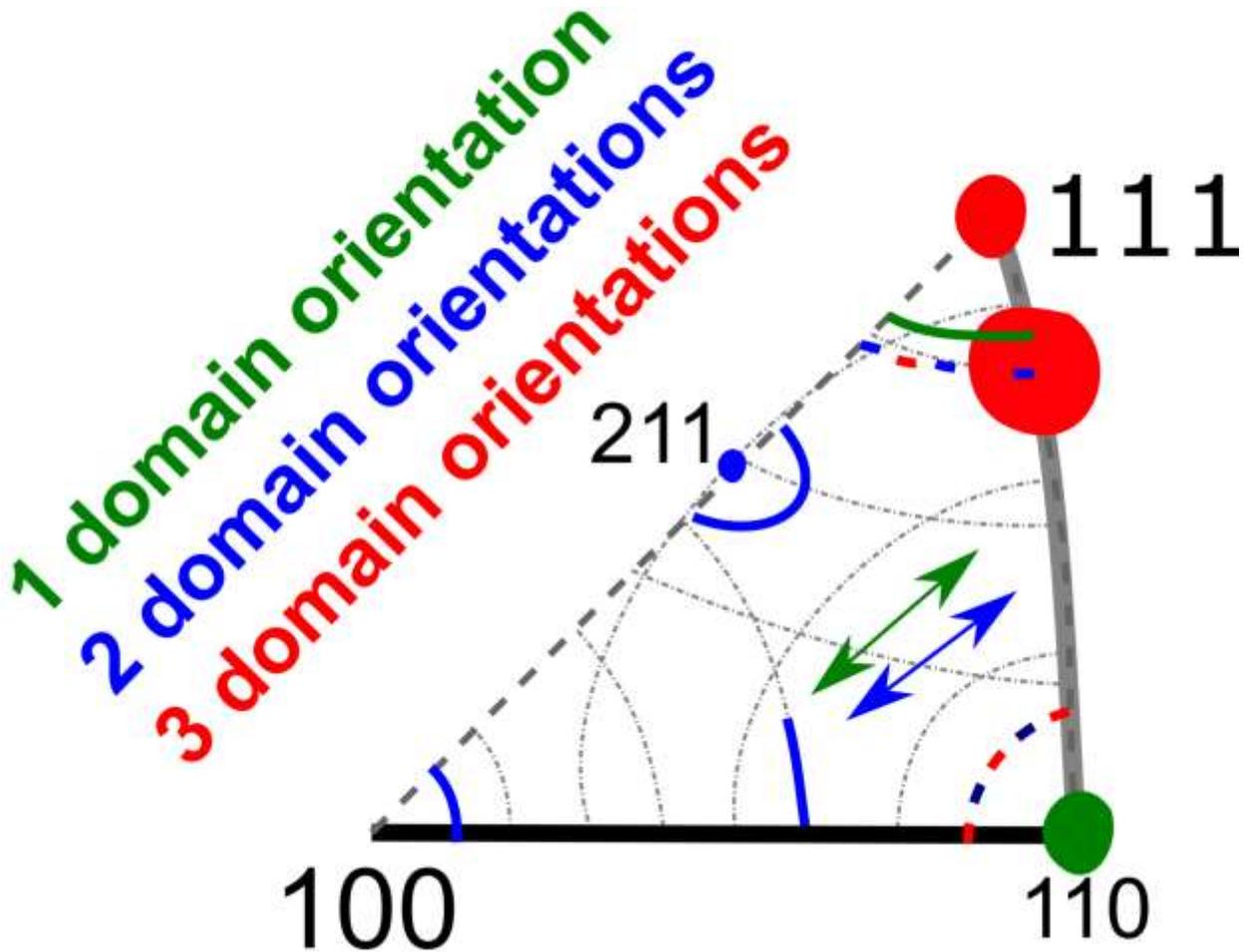

**SI-Figure 9.** Dependence of the number of domains on the TEM foil out-of-plane orientation. Due to uncertainties in the determination of the out-of-plane orientations, the range of possible normal directions for each foil are indicated by lines, arcs, or spots. The TEM foil normal is assumed here to be parallel to the TEM grid normal, although it may in fact be several degrees off. The color of the markings indicates the number of observed twin domains.

In order to look for any connections between the TEM foil out-of-plane orientations and the number of twin domains during the various in-situ tests, we summarized the observations based on the pre-characterisation of the foils using a stereographic triangle for the cubic system, **SI-Figure 9**. We have argued that the nanotwin formation is likely stress-driven. Assuming an equi-biaxial stress state in the plane of the TEM foil, the largest reduction in strain energy is predicted when (i) the c-axis of the tetragonal phase is aligned along the [100] direction of the pristine cubic phase that was closest to the foil out-of-plane normal, and (ii) the observed twin domains reflect the twinning planes with the largest resolved shear stresses. Thus, we would predict three different twin domains for foils with a near-$<111>_c$ out-of-plane orientation, two different twin domains for foils with a near-$<100>_c$ out-of-plane orientations, and one twin domain for foils with near-$<110>_c$ out-of-plane orientations. Since we did not always measure the out of plane orientation of the tetragonal phase, we are unable to test the first assumption. However, the summary of the in-situ test observations, **SI-Figure 9**, does indeed show that the number of twin domains does show a tendency to adhere to the behavior expected for microstructure formation governed by stress relief.